\documentclass[times,twocolumn,review]{elsarticle}

\usepackage{cag}
\usepackage{framed,multirow}

\usepackage{amssymb}
\usepackage{latexsym}

\usepackage{graphicx}
\usepackage{amsmath}
\usepackage{amssymb}

\usepackage{url}
\usepackage{xcolor}
\definecolor{newcolor}{rgb}{.8,.349,.1}

\usepackage{hyperref}

\usepackage[switch,pagewise]{lineno} 

\journal{Computers \& Graphics}

\begin{document}

\verso{Preprint Submitted for review}

\begin{frontmatter}

\title{Guided proceduralization: Optimizing geometry processing and grammar extraction for architectural models}%

\author[1]{\.Ilke Demir\corref{cor1}}
\cortext[cor1]{Corresponding author: 
idemir@purdue.edu}
\author[1]{Daniel G. Aliaga}

\address[1]{Purdue University, West Lafayette, IN, US}

\received{1 November 2017}
\accepted{16 May 2018}
\finalform{10 April 2018}

\begin{abstract}
We describe a guided proceduralization framework that optimizes geometry processing on architectural input models to extract target grammars. We aim to provide efficient artistic workflows by creating procedural representations from existing 3D models, where the procedural expressiveness is controlled by the user. Architectural reconstruction and modeling tasks have been handled as either time consuming manual processes or procedural generation with difficult control and artistic influence. We bridge the gap between creation and generation by converting existing manually modeled architecture to procedurally editable parametrized models, and carrying the guidance to procedural domain by letting the user define the target procedural representation. Additionally, we propose various applications of such procedural representations, including guided completion of point cloud models, controllable 3D city modeling, and other benefits of procedural modeling. 
\end{abstract}


\end{frontmatter}


\section{Introduction}

The recent popularity of 3D environments and models for augmented and virtual reality environments puts high expectations on the complexity and quality of such assets, with a desire to replicate the real world. Urban planning, remote sensing, and 3D reconstruction researchers have been focusing on bringing the digitized and physical world together, with an emphasis on urban models. In parallel to the demand for city-scale 3D urban models, the availability of 3D data acquisition systems and image-based solutions have also increased. Although using the 3D data obtained from different sources such as images, laser scans, time-of-flight cameras, and manual modeling databases is an option, the results of these approaches usually do not expose an easily modifiable model with structural parts and thus obstructs architectural reconstruction and modeling tasks. 

Aiming for automatic generation, procedural representations are highly parameterized, compact, and powerful, especially in urban domain~\cite{mathias2011,vanegas2010}. The pioneering work of Parish and Mueller~\cite{parish2001}, and subsequent urban modeling papers (e.g., see surveys~\cite{Smelik14CGF,ourcourse}) focused on forward and inverse procedural modeling approaches. While procedural modeling (PM) accomplishes providing architectural structures of required detail, creating complex realistic building templates needs time, extensive coding, and significant domain expertise. Inverse procedural modeling (IPM) addresses the shortcomings of procedural modeling by controlling and adapting the procedural generation to a given target model~\cite{martinovic2013,weissenberg2013}. In this sense, IPM can be regarded as an optimization problem over the space of derivations, to guide procedural modeling.

We want to carry this problem one step further by (1) removing the dependency on the space of derivations, and (2) switching the control domain. \textit{Proceduralization}\cite{my} takes care of the first motivation, by converting existing geometric models into a procedural representation, with no a priori knowledge about the underlying grammar. 

However, as this procedural representation aims to serve as the minimal description of the model, evaluating for the best description (e.g., the description with the best expressiveness) requires determining its Kolmogorov complexity, which is uncomputable. Our solution is to let the user guide the system to find the best grammar per use case. This also handles the second motivation by enabling the user to control the characteristics of the extracted grammar. In other words, inverse procedural modeling enhances procedural modeling by producing the best instance, while guided proceduralization enhances proceduralization by producing the best grammar.

\begin{figure}[hbt!]
	\begin{center}
\includegraphics[width=1\linewidth]{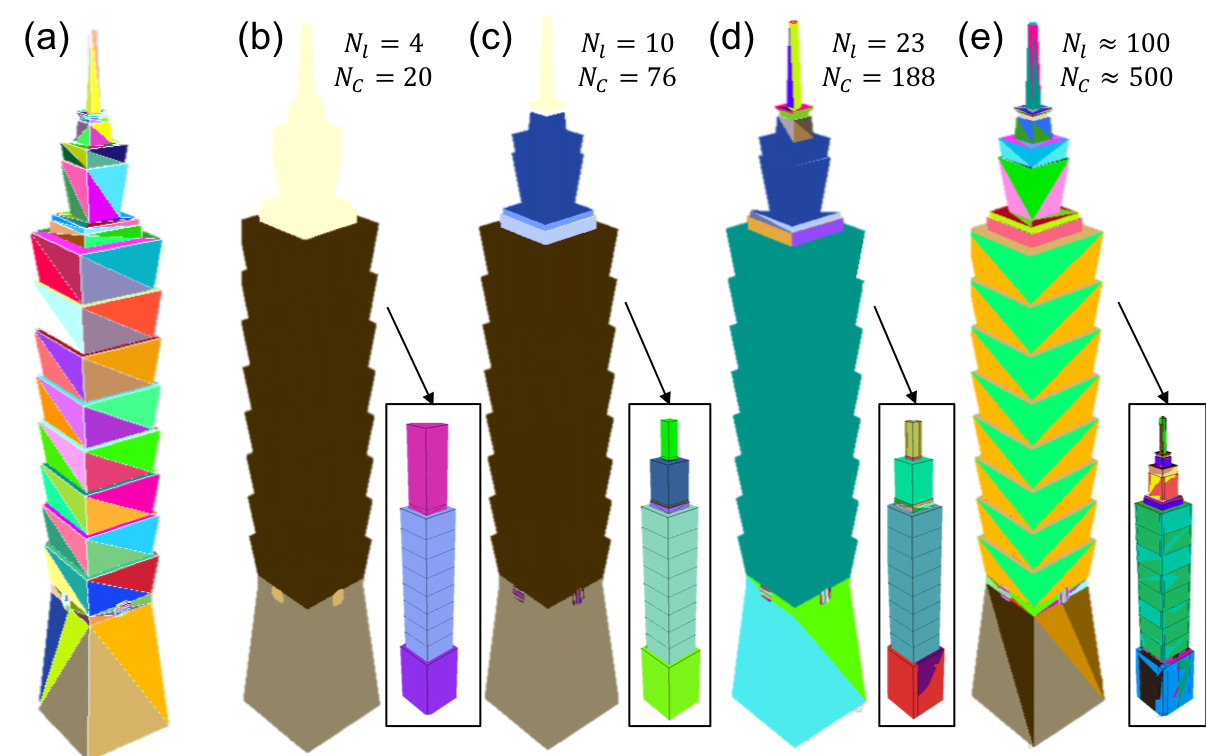}
	\end{center}
  \caption{\textbf{Guided Proceduralization.}  (a) The input model of \textit{Taipei 101}, (b-e) colored procedural elements ordered by increasing target parameter values (e.g., number of components $N_c$ and number of similarity groups $N_l$). As the user specification on the target grammar changes, different grammars of the model are revealed. Insets indicate representative instances of rules and terminals.}
  \label{fig:teaser}
\end{figure}

In this paper, we focus on guided proceduralization for 3D urban modeling and reconstruction of architectural meshes, building point clouds, and textured urban areas. Our framework provides a feedback loop which under user control seeks our the hidden high-level hierarchical and structural information coherent with the target specification and the application objectives (Figure~\ref{fig:teaser}). Although guided procedural modeling approaches (\cite{ritchie2016,benes2011}), and proceduralization methods (\cite{demir15ICCV,demir2016}) have been introduced, we propose the first approach to guide the proceduralization process using specifications of a target grammar. We start with definitions and functions for guided proceduralization, then introduce geometry processing and grammar extraction steps of generalized proceduralization in the controlled setting. Afterwards, we demonstrate applications of the obtained grammars in completion, reconstruction, synthesis, modeling, querying, simplification, and rendering.

Altogether, our main contributions include:
\vspace{-5pt}
\begin{itemize}
\item a generalized guided proceduralization framework that extracts procedural representations across different 3D data types,
\item an optimization process to evaluate and diversify the grammars output by our proceduralization,
\item a feedback loop to enable guidance to control proceduralization for obtaining the most expressive grammar and for various aims, and
\item several applications of the guided proceduralization framework for editing and merging various models.
\end{itemize}

Using our controlled proceduralization framework, we have extracted procedural models from complex polygonal models (i.e., Turning Torso, Taipei 101, Saint Basil's Cathedral), from point clouds of architectural scenes (i.e., Staatsoper Hannover, Wilhelm Busch Museum), and from textured massive city models (i.e., $180km^2$ metropolitan area of New York with more than 4000 buildings, San Francisco, Chicago). We have used these models to create more complete reconstructions, controlled procedural generation, and easier, efficient, and structure-aware urban modeling.


\section{Related Work}
\begin{figure*}[h!]
	\begin{center}

			\includegraphics[width=1\linewidth]{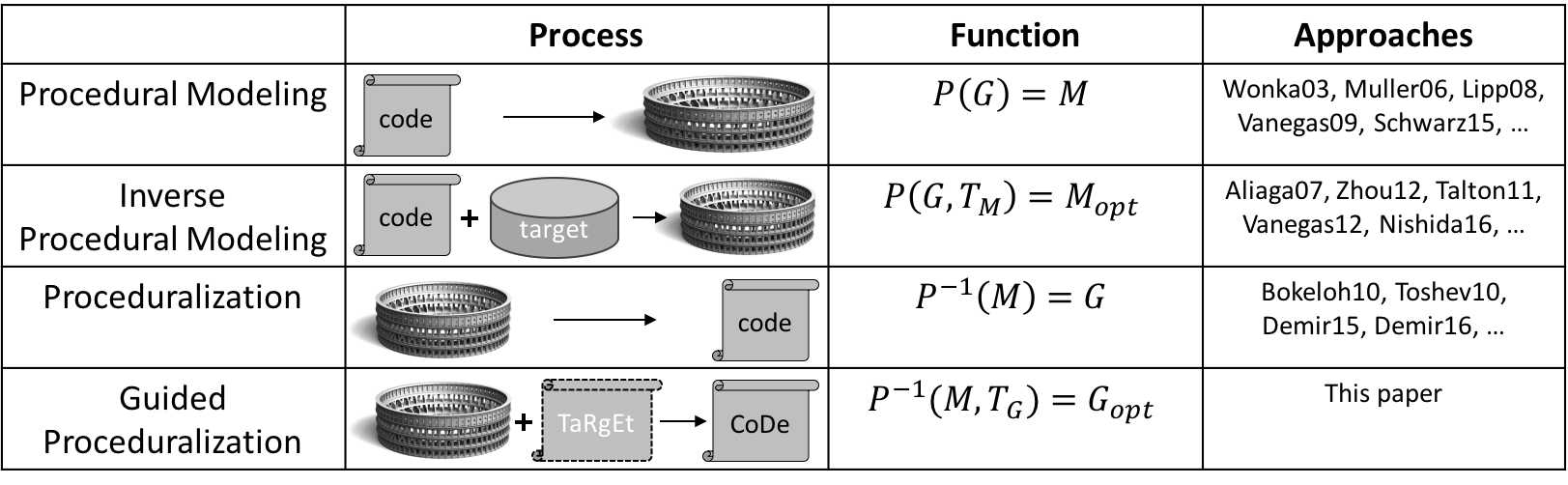}
	\end{center}
	\caption{\textbf{Evolution of Procedural Modeling.} Differences between controlled and inverse procedural modeling approaches with references. $P$ represents the process of procedural modeling with an input grammar $G$ and output geometric model $M$. $T_{M}$ indicates a geometric target instance (exemplar, silhouette, etc.), and $T_{G}$ indicates a procedural target instance (grammar elements, pattern examples, etc.). $*_{opt}$ indicates that it is optimized for ``the most representative of the target''.}
	\label{fig:prev}
\end{figure*}

\subsection{Procedural Modeling}
Procedural modeling $P$ generates a model $M$ from a given grammar $G$ ($P(G)=M$). Starting with the pattern language of Alexander et al.~\cite{patlan}, PM has been utilized by many approaches (\cite{wonka2003, muller2006, schwarz15, vanegas2009}) in urban settings. More PM approaches are surveyed by Smelik et al.~\cite{Smelik14CGF}. However, coding procedural details is a cumbersome process needing domain expertise and codification skills. Interactive editing systems, such as Lipp et al.~\cite{lipp2008} and CityEngine, have been added on top of existing procedural systems to facilitate grammar editing. These can be considered as first guidance solutions for procedural systems. 

\subsection{Inverse Procedural Modeling}
In contrast, inverse procedural modeling discovers the set of parameters, probabilities, and rules from a given grammar to generate a target instance\cite{ourcourse} ($P(G, T_M)=M_{opt}$ as per the second row of Figure~\ref{fig:prev}). Initial works provided semi-automatic and automatic building (e.g.,\cite{aliaga2007, zhou2012, toshev2010}) and facade solutions (e.g.,~\cite{bao2013, hohmann2009, musialski2013, gadde2016,teboul2013}). Given some exemplar derivations and labeled designs, Talton et al.~\cite{talton2010} use Bayesian induction to capture probabilistic grammars. Similarly, Monte Carlo Markov Chain (MCMC) optimization is used to discover the optimized parameters for a target instance of a procedural representation of buildings (Talton et al.~\cite{talton2011} and Nishida et al.~\cite{nishida2016}) and cities (Vanegas et al. \cite{vanegas2012}). Most of those solutions support synthesizing similar models that best fit the given guidance. However they rely on pre-segmented components, known grammars, and known layouts to generate the derivation space. This is an important drawback, since it constrains reconstruction and modeling to the limited space of the initial grammar. In contrast, we want to carry the guidance from the geometric space to the procedural space, thus the \textit{desired control} is defined rather than the \textit{desired model}.

\subsection{Proceduralization}
Proceduralization starts with only geometry and no knowledge on the grammar ($P^{-1}(M)=G$ as per the third row of Figure~\ref{fig:prev}). Some image based techniques use deep learning for extracting simple and-or template grammars~\cite{si2013}, or generative schemes~\cite{demir2017}. In urban scenes, Bokeloh et al.~\cite{bokeloh2010} use partially symmetric structures to search for transformations that map one partition to another based on $r$-similar surfaces. It enables building model synthesis though not formally yielding a procedural model. For point clouds, Toshev et al.~\cite{toshev2010} segment a building into planar parts and join them using a hierarchical representation that separates roof, rest of the building, and non-building structures. Demir et al.~\cite{demir15ICCV} focus on point clouds, and user control is explicit in the geometric domain at the semi-automatic segmentation step. Martinovic et al.~\cite{martinovic2013} use Bayesian induction to obtain facade grammars, and Kalojanov et al. \cite{kalojanov2012} divide the input structure into microtiles to detect partial similarities. Demir et al.~\cite{demir2016} introduces proceduralization, automatically creating a set of terminals, non-terminals, and rules (blue path in Figure~\ref{fig:pipeline}). Although evaluating the expressiveness of this automatic encoding is uncomputable (see Section~\ref{sec:expressive}), the expressiveness of a fixed grammar per model is still limited with regard to the modeler's use case. Thus, their approach has smaller tolerance for noise, directed by the one-pass segmentation and labeling, is not flexible for different use-cases, works only on triangular models, and does not allow user control (for the grammar generation).

\subsection{Guided Proceduralization}
\begin{figure}[b!]
	\begin{center}

			\includegraphics[width=1\columnwidth]{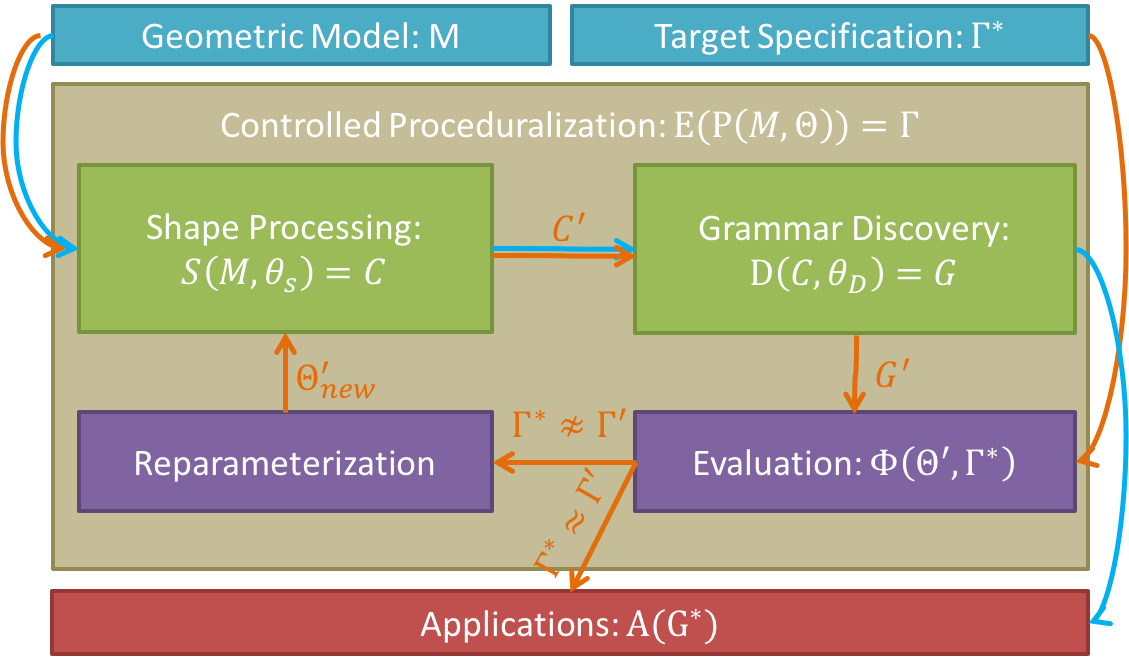}
	\end{center}
	\caption{\textbf{Pipeline.} Our system optimizes parameter values $\Theta$ of the proceduralization function $P$ to generate a grammar $G$ that fulfills user-provided grammar values $\Gamma^*$. Blue path shows traditional proceduralization pipeline and orange path shows guided proceduralization pipeline.}
	\label{fig:pipeline}
\end{figure}

In contrast, the key motivation behind our research is that, if we have some insights about the desired grammar, we can evaluate the proceduralization outputs to suggest candidate grammars. This creates a pioneering framework that is the first to provide guided proceduralization for synthesis of arbitrary 3D architectural structures ($P^{-1}(M, T)=G$ as per the last row of Figure~\ref{fig:prev}). As guidance in procedural modeling enables finding \textit{the best instance}, guidance in proceduralization enables finding \textit{the best grammar}. This guidance enables the procedural representation to be optimized by user specification (orange path in Figure~\ref{fig:pipeline}), so that the resulting grammars are robust to model noise, flexible for different use cases, independent of segmentation and labeling, and supports more creative processes. In summary, we would like the artists to define the desired control, instead of the desired structure, while still being able to derive procedural representations from existing models.

\section{Definitions and Functions for Guidance}

\begin{figure}[hb!]
	\begin{center}

			\includegraphics[width=1\columnwidth]{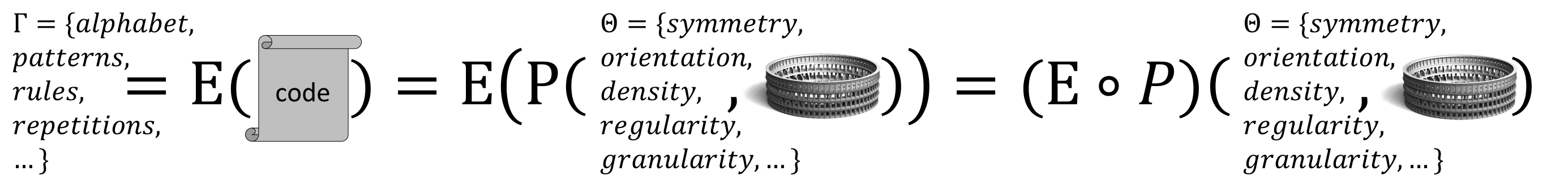}
	\end{center}
	\caption{\textbf{Evaluation of Proceduralization.} The input model is proceduralized using some input parameter set $\Theta$. The resulting grammar is evaluated to obtain grammar specification values $\Gamma$.}
	\label{fig:pipe2}
\end{figure}

Our control loop optimizes the internal parameters of the proceduralization process to generate a procedural description (e.g., a grammar) that meets a user-provided set of grammar specification values. More precisely, proceduralization function $P$ operates on a set $\Theta =\{\theta_1,\theta_2,\dots,\theta_m\}$ of $m$ input parameter values and an input model $M$ to produce a grammar $G$. The grammar is then assessed by an evaluation function $E$ which produces a set $\Gamma=\{\gamma_1,\gamma_2,\dots,\gamma_n\}$ of $n$ grammar specification values (or simply grammar values). Each parameter value and grammar value should be within a range of $[\theta_{m_{min}},\theta_{m_{max}}]$ and $[\gamma_{n_{min}},\gamma_{n_{max}}]$, respectively. Symbolically, we can define the relationship of input parameters to the target grammar values as follows in Eqn.1, which is also depicted in Figure~\ref{fig:pipe2}.
\begin{equation}
\Gamma=E(G)=E(P(\Theta, M))=(E\circ P)(\Theta, M)
\end{equation}

Intuitively, guided proceduralization frees grammar discovery from any dependency to the underlying segmentation and grammar discovery. There is no concept of "good segmentation", "different tessellations", "noisy point clouds", "vertex displacements", because the whole idea of this paper is to let the optimization find the best internal parameters towards the target grammar specification. Previous proceduralization approaches failed to achieve that, because evaluating grammars for expressiveness is uncomputable. However, because we are provided with an estimate of "the best grammar" by the user, we keep optimizing the whole proceduralization process (including segmentation, labeling, pattern analysis, etc.) to converge for the specification.

\subsection{Target Grammar Specification}
\begin{figure}[hb!]
	\begin{center}
			\includegraphics[width=1\columnwidth]{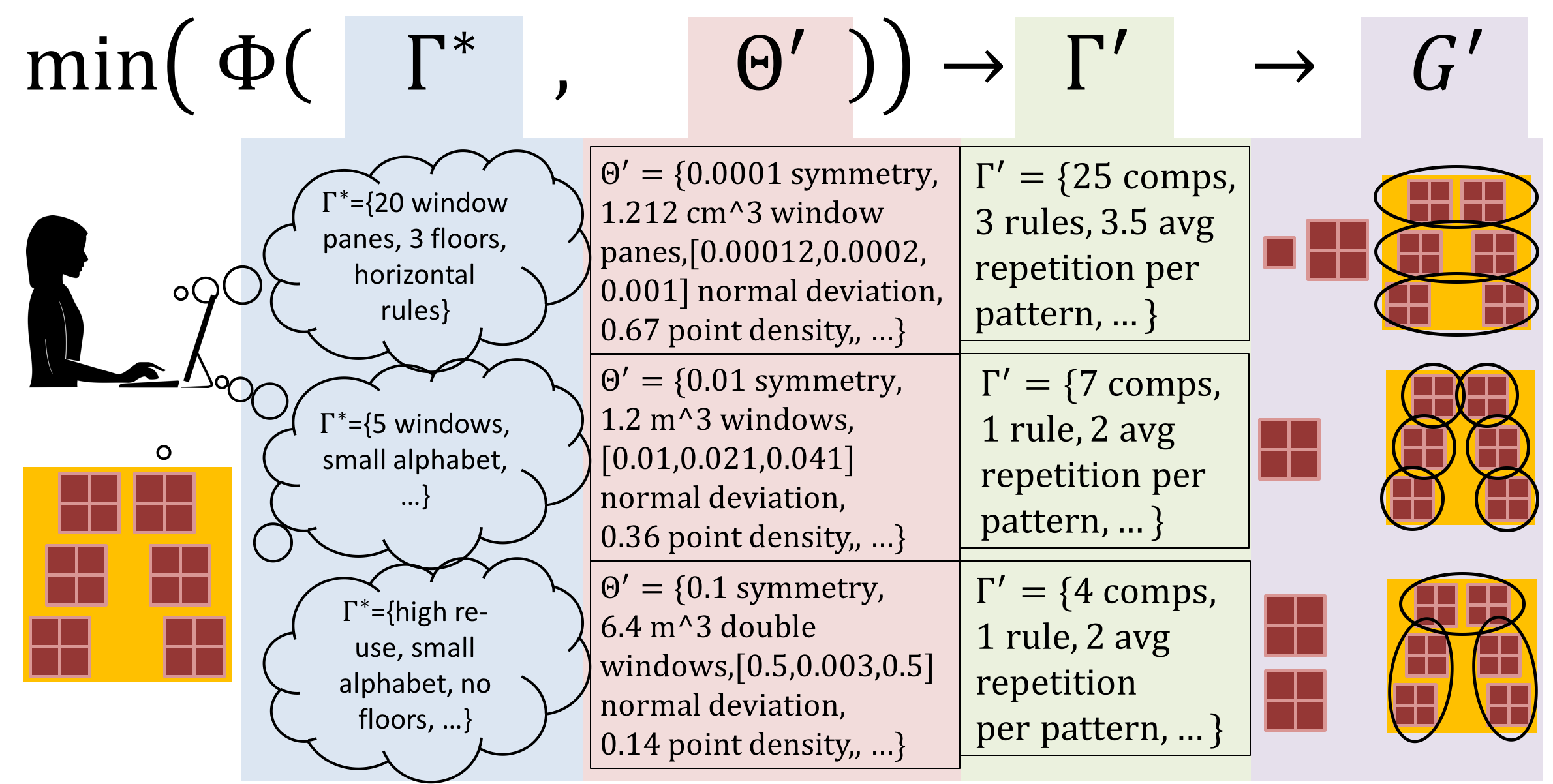}
	\end{center}
	\caption{\textbf{Optimizing for Target Grammars.} Minimization of the error function $\Phi$, finds the optimum set of input parameters $\Theta'$ that generates $\Gamma'$ which is the closest grammar specification to the user defined set $\Gamma^*$. As depicted, each $\Gamma^*$ ends up generating a different grammar $G'$, unlike traditional proceduralization. The first row has window panes as the terminals, windows as rules. The second row has a window as a terminal, and the last row has a double window with varying orientation, as a terminal. Example rules per grammar are also given per input $\Gamma^*$.}
	\label{fig:min}
\end{figure}

Our optimization objective is to let the user specify the desired grammar values and then to automatically generate a grammar whose actual grammar values are close to the specified ones. Hence, we define the set $\Gamma^*=\{\gamma^*_1,\gamma^*_2, \dots,\gamma^*_n\}$ to be the user-specified target grammar values. Then, our 
method computes a set of parameter values $\Theta'$ that leads the proceduralization function $P$ to produce a grammar $G'$ exhibiting grammar values closest to the target values $\Gamma^*$. In other words, we seek a set of parameter values $\Theta'$ such that $(E\circ P)(\Theta', M)\rightarrow \Gamma^*$. The equality is replaced by an arrow to indicate that we want the convergence, as absolute equality may not be satisfied for every case. The overall error is encapsulated in the function $\Phi(\Theta', \Gamma^*)$ and the computational goal is to find values of $\Theta'$ that minimizes $\Phi$. Figure~\ref{fig:min} gives three examples for a toy facade model, where the user defines different grammar values to expose different grammars.

\subsection{Error Function Approximation} 
Our control loop requires calculating $(E\circ P) (\Theta, M)$ for a large number of different values for $\Theta$. Prior inverse procedural modeling and early proceduralization systems use predefined numbers for $\Theta$ and typically are not set up to generate such a large number of productions at reasonable rates. Hence, we simplify the $(E\circ P)$ by a function $f(\Theta)$ that does not output the grammar but generates the set $\hat{\Gamma}=\{\hat{\gamma_1},\dots,\hat{\gamma_n}\}$ that approximate the grammar values. The function $f$ is a good approximation of $(E\circ P)$ if $|\Gamma-\hat{\Gamma}| < \epsilon$ where $\epsilon$ is a small positive constant, meaning that each grammar value $\hat{\Gamma_n}$ generated by the approximator should be close enough to the corresponding actual grammar value $\Gamma_n$ generated by forward proceduralization. Thus, given a good approximation $f$ of the proceduralization system and grammar evaluation $(E\circ P)$, our optimization goal is to find sets of parameter values $\Theta'$ such that $f(\Theta')\rightarrow \Gamma^*$.

\section{Guided Proceduralization Framework}
In the proceduralization framework $G=P(M)$ (Figure~\ref{fig:pipeline}, blue path), the aim is to obtain the underlying grammar $G$ that is representative of the input model $M$. The input to this pipeline can be images, meshes, point clouds, and almost any content that can be categorized. The granularity of the content for urban proceduralization can vary from facades, to buildings, and to entire cities. Also, the input can be some specific elements of an urban area such as parcels
, roads
, buildings
, or a full city
. The proceduralization pipeline is divided into two main steps: geometry processing and grammar discovery. 

For guided proceduralization (Figure~\ref{fig:pipeline}, orange loop), the proceduralization function is parametrized such that different grammars can be extracted based on different parameters (i.e., $G=P(\Theta,M)$). Those grammars do not only differ by the composition of the rules and rule parameters, but the terminals and non-terminals also vary between different grammars, as the decomposition changes. The parametrization is carried over to the two main stages of proceduralization. Shape processing $S(\Theta_S, M)=C$ in turn is subdivided into segmentation and labeling, while grammar discovery $D(\Theta_D, C)=G$ is further separated into hierarchy construction and pattern extraction. We can re-write the proceduralization framework in detail as follows:

\begin{align}
G&=P(\Theta,M)=D(\Theta_D,C) \nonumber\\
&=D(\Theta_D,S(\Theta_S,M))\nonumber\\
&=(D\circ S)(\Theta,M)
\end{align}

\subsection{Geometry Processing}
Our geometry processing step $S(\Theta_S,M)$ detects the repetitive nature in the underlying buildings, and uses that to compute a segmentation and perform a similarity labeling. The input model $M$ will be segmented into components $c_i$, which are the smallest geometric units of decomposition, also candidates to become terminals of the grammar. Each component will have a corresponding label $l_j$, where the label is the indicator of the similarity group that the component belongs to. Thus, the full set of components is $C=\{(c_1,l_1),\dots ,(c_{N_c}, l_{N_l})\}$ (where $N_l \leq N_c$). Using the control loop, different $C$s are produced depending on the $\Gamma^*$ values (notice the different set of components and patterns found for the middle section of the tower in the top right corner of Figure~\ref{fig:difgram}).
\begin{figure}[hbt!]
	\begin{center}
			\includegraphics[width=1\columnwidth]{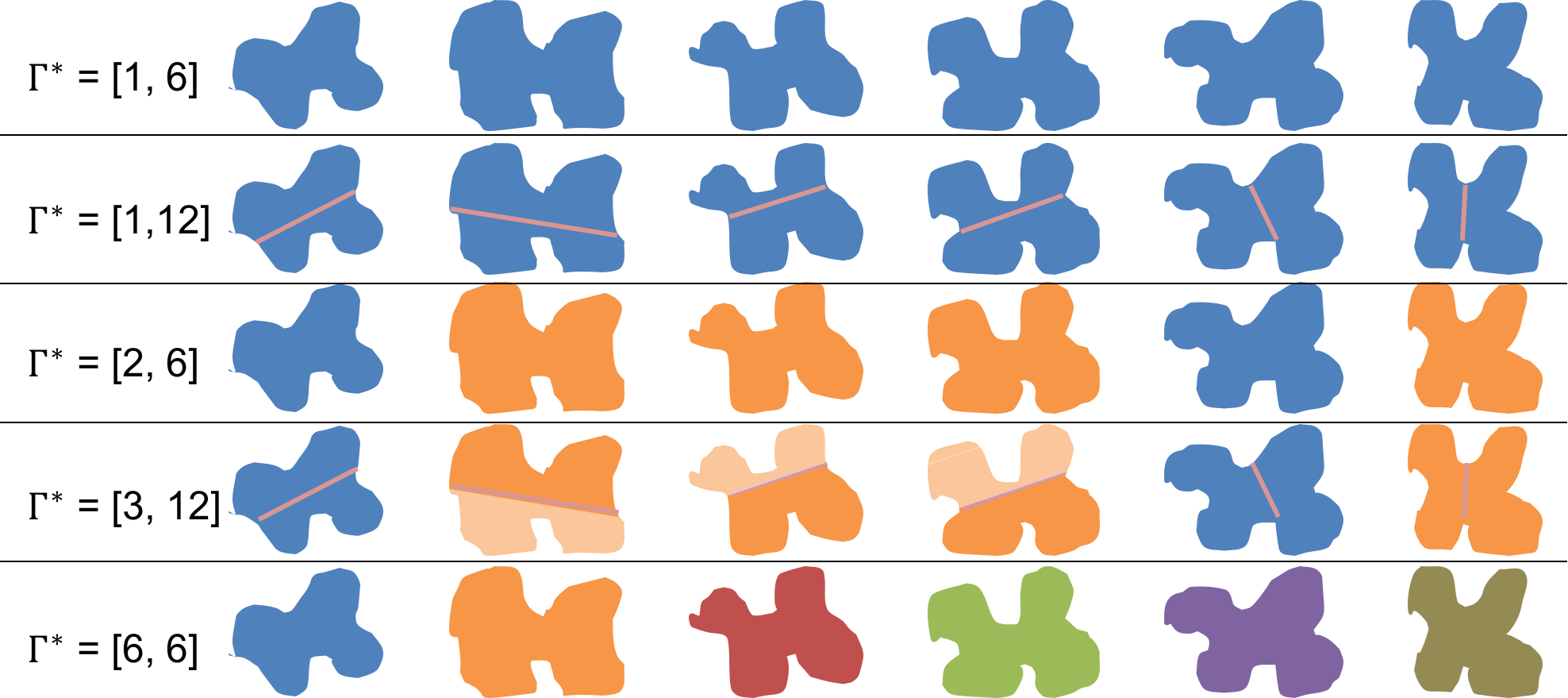}
	\end{center}
	\caption{\textbf{Shape Processing.} The effect of reparameterization of $\Theta'$ based on user-specified grammar values $\Gamma^*$ is illustrated on the segmentation on a set of components with six similar but irregular shapes. The colors indicate the similarity groups $l_j$. As an example, the target grammar values for row 4 are ``3 component types and 12 components'', ending up segmenting the set to 5 orange, 4 blue, and 3 light orange components. }
	\label{fig:shape}
\end{figure}

\subsubsection{Parametrized Shape Segmentation}
Our formulation for $S$ depends on the type of input data $M=\{t_1,\dots , t_{N_m}\}$. Prior segmentation/labeling work focuses on single data types and exploiting geometric similarities \cite{lipman2010}, analogies with the exact-cover problem~\cite{demir2015}, image contours \cite{zhou2012}, pre-defined semantics~\cite{toshev2010, hohmann2009}, or geometric primitives~\cite{attene2006,mathias2011}. The aforementioned approaches do not organize the repetitions into patterns, use the patterns for modeling, or provide a way to control pattern granularity selection. 

In contrast, our shape processing method extends prior work by adapting it to our guided proceduralization framework, by introducing weights as parameters to balance collective component properties such as number of components and number of similarity groups. As shown in Figure~\ref{fig:shape}, various (simplified) grammar values end up generating several versions of segmentation and labeling of the same set. Here, $\Gamma^*$ only consists of the alphabet size and the component size, and the components are color-coded by their labels.

\subsubsection{Segmentation Formulation}\label{sec:thetas}
Our approach combines segmentation and labeling in a coupled algorithm, where the model is regarded as the combination of all elements of all labels (Eqn.~\ref{eqn:model}), and shape processing outputs the components and their labels (Eqn.~\ref{eqn:sha}).
\begin{align}
M&=\Sigma_i^{|M|}{t_i}=\Sigma_j^{N_l}{\Sigma_k^{N_c}{(t_i,l_j)\ |\ t_i \in c_k}}\label{eqn:model}\\
S(M,\Theta_S)&=\{(\Sigma_k^{|c_1|}{t_k},l_1),\dots ,(\Sigma_k^{|c_{N_c}|}{t_k},l_{N_l})\}\label{eqn:sha}
\end{align}
Our parameter value vector contains similarity-metrics for the components, namely $\Theta_S=\{\theta_{geo}, \theta_{top}, \theta_{den}, \theta_{dir}, \theta_{num}\} $ for geometric similarity, topological similarity, density, orientation, and number of elements, respectively. 

\begin{figure*}[htb!]
	\begin{center}
			\includegraphics[width=1\linewidth]{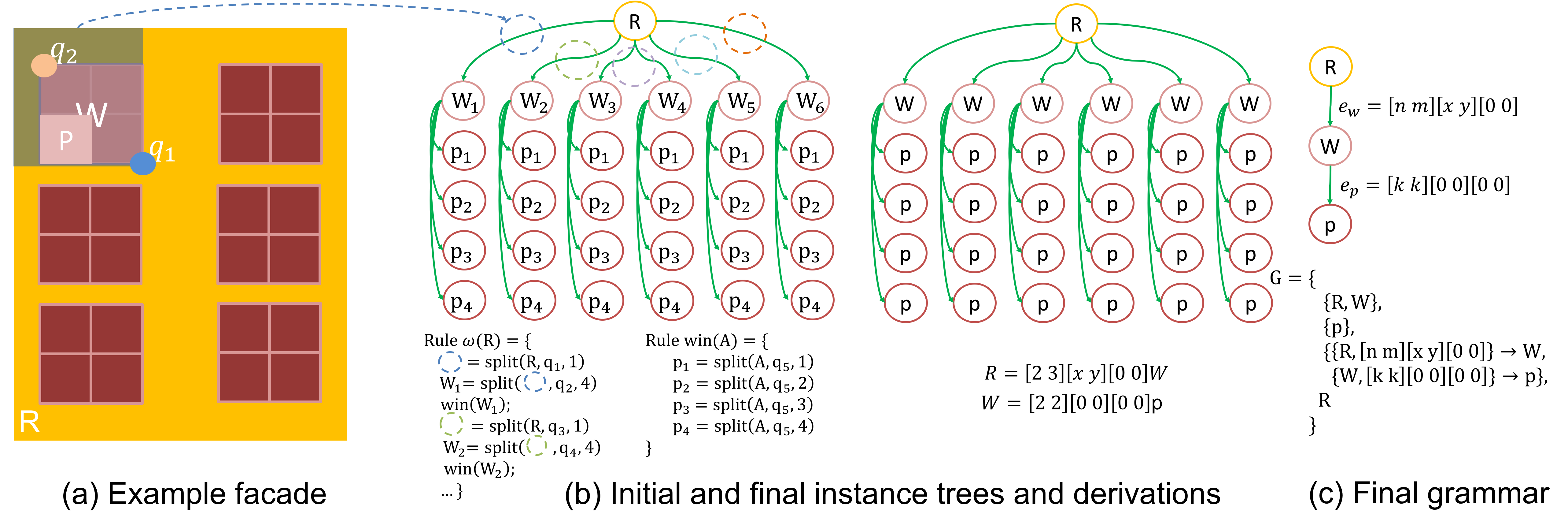}
	\end{center}
	\caption{\textbf{Split Tree and Grammar.} (a) Components of the example facade: $p$ as a windowpane and $W$ as a window. $q_1$ is the split point to create the invisible blue node, and $q_2$ is the second split to obtain $w$ from the blue node. (b) initial and final split tree and corresponding rules: $p$ recognized as a terminal and $W$ recognized as a non-terminal, meanwhile rule parameters are discovered, and (c) the final grammar with a facade rule and a window rule.}
	\label{fig:gramex}
\end{figure*}

\subsubsection{Input Dependent Implementation}\label{sec:inpdep}
Although the metric is unified into one representation, the segmentation method and the elements in the parameter value vector varies based on the data type of $t_i$. Our framework collectively optimizes over these methods, where traditionally hand-tuned parameters are replaced by optimization space parameters specific to the input type. In all cases, we will use our overall optimization framework to compute the parameter values $\Theta'_S$ that best yields a grammar with the desired values $\Gamma^*$, expressed as $(E\circ (D\circ S))(\Theta'_S,\Theta'_D, M)\rightarrow \Gamma^*$.

\textit{4.1.3.1. Triangular Models}

For example, for $t_i$ as triangles, the similarity parameter $\theta_{geo}$ is defined as a weighted sum of similarity of convex-hulls and averaged similarity of triangles between two components. $\theta_{num}$ becomes number of initial component seeds for triangles. $\theta_{dir}$ becomes the similarity of orientations by a deviation of normal vectors. The segmentation methodology also changes based on the data type. For triangular meshes, we use synchronous volume growing~\cite{demir2015} to establish a segmentation and similarity detection simultaneously, where the algorithm parameters are explored by the optimization implicitly.

\textit{4.1.3.2. Point Clouds}

For $t_i$ as points, geometric similarity metric $\theta_{geo}$ is defined as the mean of correspondence distances normalized by point density, after an iterative closest point (ICP)\cite{icp} alignment between components, because correspondence distances without aligning the two components result in incorrect measurements for $\theta_{geo}$. For the segmentation of point clouds, we use a loop of plane fitting and Euclidian clustering ~\cite{rusu2011}, where $\theta_{num}$ represents the iteration count for the expected number of segments. $\theta_{den}$ defines the minimum density of components, which is also instrumental in cleaning the point clouds during segmentation phase.

\textit{4.1.3.3. Textured Meshes}

Lastly, for $t_i$ as textured polygons, $\theta_{geo}$ is defined as a similarity vector of components with multiple geometric and visual features (dimensions, albedo, color patterns, etc.). The segmentation incorporates rendering components in different main directions and measuring the rate of change for finding the cut planes at significant jumps. The components between the cut planes are re-rendered for extracting and clustering their feature vectors~\cite{demir2014}. In this process, $\theta_{num}$ controls the camera displacement step size. 

\subsection{Grammar Discovery}
The second step of the pipeline organizes the segments $c_i$ and their labels $l_j$ into a hierarchy, which we will call an instance tree $T=\{(v_0, \varnothing), (v_1, e_1), \dots, (v_{N_v}, e_{N_v})\}$, where $v_i$ denotes a tree node (with $v_0$ being the root), and $e_i$ denotes the incoming edge to node $v_i$. This hierarchy is needed to encode the relations of $c_i$ and to discover the underlying rules. It also encodes an actual derivation instance thus can be treated as an early procedural representation of the model. After the hierarchical structure is constructed, it is analyzed for patterns and for the distribution of the components to finally extract the grammar $G$.

\subsubsection{Grammar Definition}\label{sec:gramval}
While $G$ can be a shape grammar or an L-system, we define a split grammar of the following form:
\begin{align}
G=&\{N, \Sigma, R,\omega\},\ where \\
N=&\{v_i\ |\ (v_i, \_)\in T\ \wedge \ \mathit{fanout}(v_i)\neq 0 \wedge (v_i,\_) \in C\}\nonumber\\
\Sigma=&\{v_i\ |\ (v_i, \_)\in T\ \wedge \mathit{fanout}(v_i)=0 \wedge (v_i,\_) \in C\}\nonumber\\
R=&\{(v_i, e_j ,v_j)\ |\ (v_j ,e_j)\in T\wedge v_i \in \mathit{anc}(v_j) \wedge \{(v_i,\_), (v_j,\_)\} \subset C\}\nonumber\\
\omega=&\{v_0\ |\ (v_0,\varnothing) \in T \wedge (v_0,\_) \in C\nonumber \}
\end{align}
where $N$ is the set of non-terminals (root nodes of subtrees in $T$ each representing a rule), $\Sigma$ is the set of terminals (leaves), $R$ is the collection of rules containing triplets of the application node $v_i$, the actual rule on the edges $e_j$ and the produced node $v_j$ (e.g., all split operations encoded in the edges $e_j$ from a subtree root node $v_i$), $\omega$ is the starting axiom (root node), $\mathit{fanout}(v_i)$ is the number of edges from node $v_i$ to its children, $v_x \in \mathit{anc}(v_i)$ means node $v_x$ is an ancestor of node $v_i$, and $\_$ denotes a ``don't care'' edge, node, or label. 

Figure~\ref{fig:gramex} depicts the grammar elements and some relations between the tree, the instance, and the grammar. The example facade in (a) contains six windows and four windowpanes per window. During the initial tree construction, the windows are produced from the wall (root) by two split operations. For the first window, R is split from $q_1$ to obtain the blue intermediate node, and the blue node is split from $q_2$ to obtain the window node$W_1$. Having no offset from the window, each window pane is produced from the window with one split operation. Leaf nodes become candidates for $\Sigma$, and subtree roots become candidates for $N$, because they mostly contain geometry. Intermediate nodes are embedded as individual split operations as a part of a rule. After we process the instance tree, we converge on the pattern structure to be discovered as the second subtree in (b), with window  pattern parameters of "2 horizontal repetition with x spacing and 3 vertical repetition with y spacing" and windowpane parameters of "2 horizontal and 2 vertical repetitions with 0 spacing". Then those pattern parameters are generalized and exported as the grammar following the syntax introduced above.

We define the grammar specification values as $\Gamma=\{\gamma_{alp},\gamma_{non},\gamma_{fan}, \gamma_{rep}\}$ which are alphabet size $|\Sigma|$, number of non-terminals $|N|$, average repetitions per pattern $\bar{|R_i|}$, and average number of components $|C|$, respectively.

\subsubsection{Tree Definition}
The components in the set $C$ of the previous section are represented by their bounding boxes $\mathit{BB}(c_i)$ and all elements $t_k \in c_i$ (e.g., triangles, points, textured polygons, etc.) are placed relative to the coordinate space defined by $\mathit{BB}(c_i)$. We start with putting all $c_i$ into an instance tree $T$ based on volumetric containment (Eqn.~\ref{eqn:bb}), based on point splits. The components are processed by decreasing size, and the tightest containing node is selected as the parent at each step. Thus, a component can only be completely inside one component, preventing $T$ from becoming a DAG.  Each $c_i$ corresponds to a node $v_i$ and each edge $e_i$ corresponds to the spatial transformation matrix (translation and rotation) of node $v_i$ relative to its parent $\alpha(v_i)$ (Eqn.~\ref{eqn:ed}): in this notation $e_i$ is a matrix and $\mathit{BB}(v_i)$ is represented by two vectors containing the minimum and maximum points of the bounding box, thus the multiplication operator gives the transformed bounding box for the new node. The labels are also preserved (Eqn.~\ref{eqn:lab}) at the initiation of $T$. $L(v_i)$ operation retrieves the label of the component $c_i$ corresponding to $v_i$.
\begin{align}
v_j=\alpha(v_i) &\iff \mathit{BB}(v_i)\subset \mathit{BB}(v_j)\label{eqn:bb}\\
(v_i,e_i) \in T &\implies \mathit{BB}(v_i)=e_i*\mathit{BB}(\alpha(v_i)\label{eqn:ed})\\
L(v_i)=l_j &\iff (v_i, l_j)\in C\label{eqn:lab}
\end{align}

\begin{figure}[ht]
	\begin{center}
			\includegraphics[width=1\columnwidth]{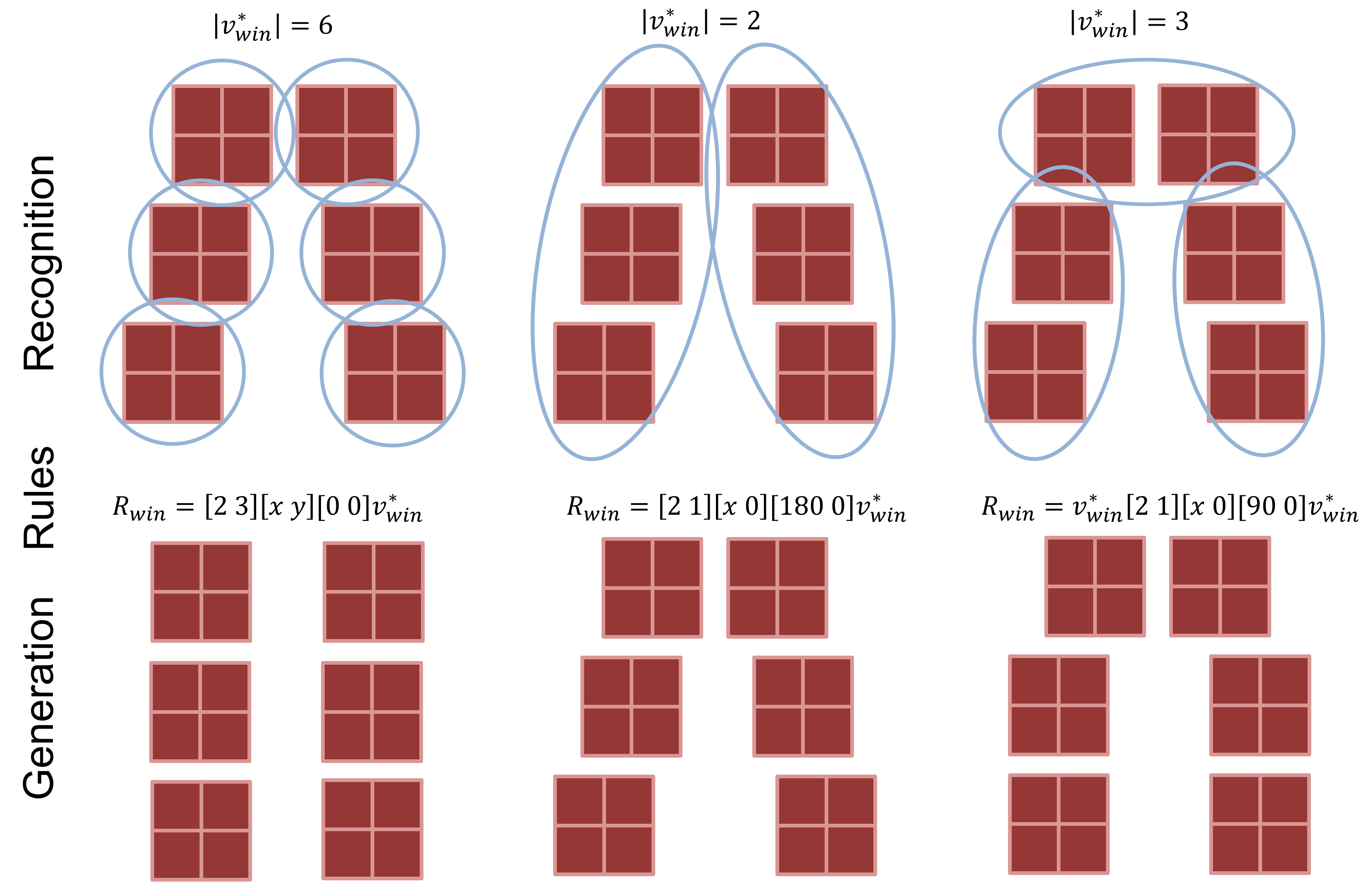}
	\end{center}
	\caption{\textbf{Grammar Discovery.} Granularity of recognized rules $R$ changes based on specified grammar values $\Gamma^*$. Extracted rules generate different derivations. The rule format is $R=[repetition][spacing][rotation][representative instance]$.}
	\label{fig:gram}
\end{figure}
\subsubsection{Subtree Similarity}
Afterwards we process $T$ to detect similar subtrees. Starting with initial labels, we compare all $v_i, v_j$ pairs where $L(v_i)=L(v_j)$. The comparison is based on pairwise topological ($\theta_{sub}$ = subtree matching by BFS ordering) and geometrical ($\theta_{sym}$ = label matching) similarity, with the additional comparisons of number of elements ($\theta_{ele}$) and size of bounding boxes ($\theta_{box}$). According to these metrics, if $v_i$ and $v_j$ are not similar, we update their labels to reflect that $L(v_i)\neq L(v_j)$. We define the concept of representative instance $v^*_x$, which indicates the common representation for a repeating subtree, for all similar nodes ($v^*_x=\{v_i\ |\ \forall i,\ L(v_i)=x\}$). Note that representative instances are not decided based on initial labels, so Eqn.~\ref{eqn:lab} is only used during hierarchy construction phase. We also perform a special \textit{join} operation to canonicalize the subtree. The derivation of nodes $v_i \in v^*_x \wedge v_j \in v^*_x \wedge \exists k |\ v_k \in \mathit{anc}(v_i) \wedge v_k \in \mathit{anc}(v_j)$ are unified to follow the same subtree structure, so that they have the same ancestor $v_k$ as the application subtree root. In this case, $v_k$ can be the parent or grandparent of joint nodes in $v^*_x$, but we will refer to it as $\alpha(v^*_x)$ to keep the notation simple.

\begin{figure*}[h!]
	\begin{center}
			\includegraphics[width=1\linewidth]{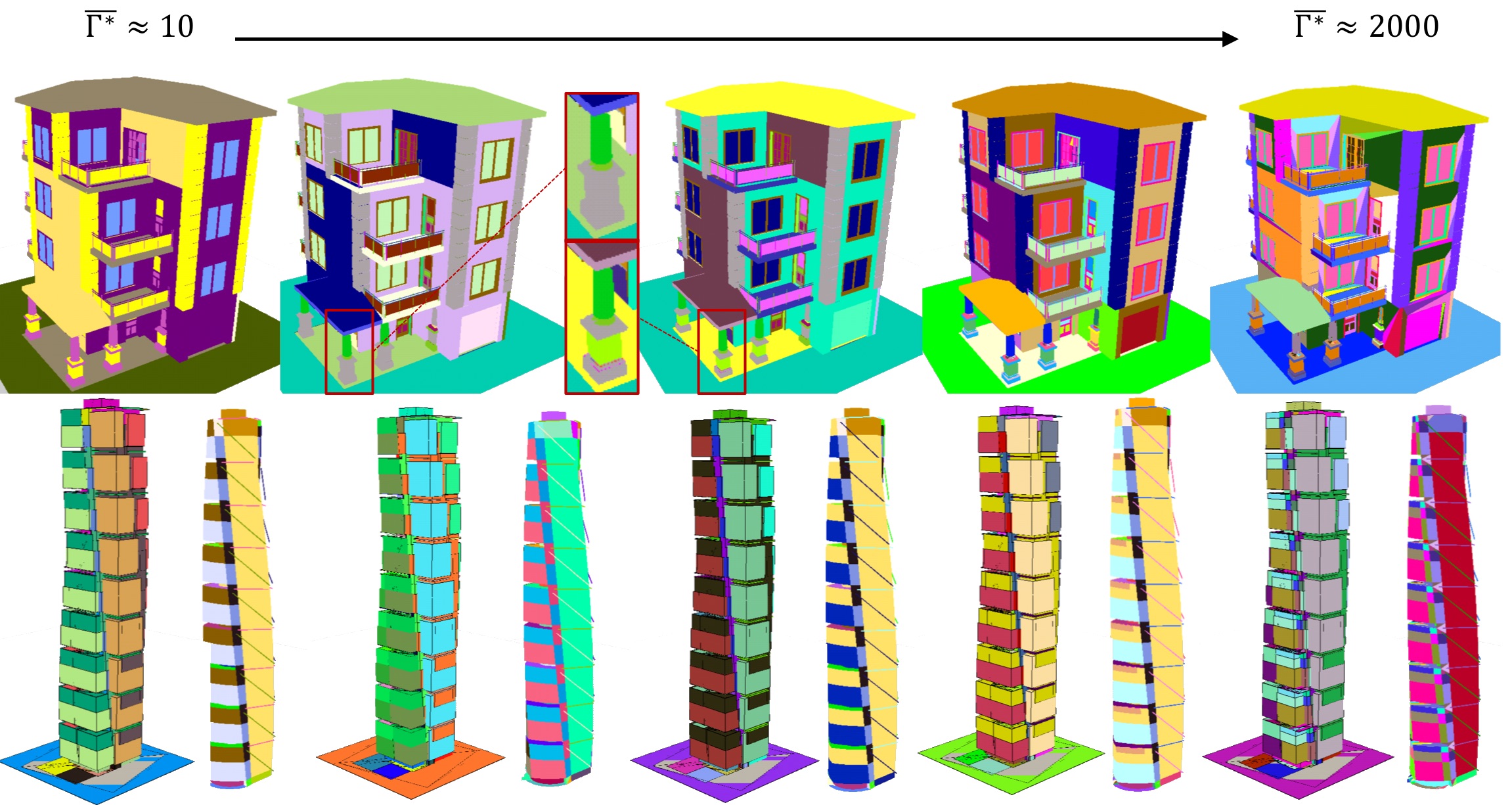}
	\end{center}
	\caption{\textbf{Varying Grammar Values $\Gamma^*$.} Grammar specification varies from small (left) to larger values (right) producing different components, labels, rules, and grammars. As can be observed in the inset of the top row, even the slight changes in specification can converge to a significantly different set of parameter values $\Theta'$. Bottom row shows a similar analysis on \textit{Turning Torso}, each instance (left) paired with its node distribution (right).}
	\label{fig:analysis}
\end{figure*}
\subsubsection{Pattern Parameters}
After all $v^*_x$ are discovered, we reveal the distribution of the representative instances and the spatial parameters of the distributions \--- namely the patterns. $R$ corresponds to an encoding of the patterns, however instead of representing each derivation with one rule on one selected edge (i.e., $R_x=(\alpha(v^*_x), e_i, v_i)\ |\ \exists \ i, \ v_i \in v^*_x \wedge L(v_i)=x$), the derivation of representative instances are combined under an averaged rule (i.e., $R_x=\{\alpha(v^*_x), \bar{e_i}, \bar{v_i})\ |\ \forall v_i \in v^*_x,\ \bar{v_i}=\mathit{mean}(v_i) \ \wedge \ \bar{e_i}=\mathit{mean}(e_i)\}$. In other words, similar derivations in the parse tree are averaged or combined to obtain a representative rule that creates the example derivation. As mentioned in the previous section, this merging process is controlled by the feedback loop, thus we introduce two more parameter values as $\theta_{pat}$ for the pattern granularity and $\theta_{ins}$ for instance similarity. The first one controls pattern granularity by adjusting the repetition count per instance and the second one controls similarity of transformation matrices to be averaged. Together, they allow the same geometrical pattern to be recognized into different rules (Figure~\ref{fig:gram}).

\subsubsection{Grammar Discovery Formulation}\label{sec:thetad}
The full parameter value vector contains similarity metric parameters for the tree nodes and for the patterns, namely $\Theta_D=\{\theta_{sub},\theta_{sym},\theta_{ele},\theta_{box}, \theta_{pat}, \theta_{inst}\}$. The control loop executes the evaluation function $E$ to determine the approximated values $\Gamma'$ and the error function $\Phi$ decides whether $\Gamma'\approx \Gamma^*$. Thus, the loop explores for the best representative instance in the optimization stage, again solving for the closest parameter values $\Theta'_D$ for the desired grammar values $\Gamma^*$, expressed as $(E\circ (D\circ S))(\Theta'_S,\Theta'_D)\rightarrow \Gamma^*$. Also, instead of running the whole framework for each optimization step, we simplified $(E\circ P)$ with $f$, which performs a reduced set of segmentation and hierarchy building operations (i.e., instead of creating a tree, numerically deciding the relationships), still yielding the same grammar values without processing the entire geometry.

\subsubsection{Grammar Output}
Finally, $G$ is exported using a common split grammar syntax where leaf representative instances are converted to terminals $\Sigma$ , and non-leaf representative instances are converted to non-terminals $N$. The root of the instance tree is output as the starting axiom $\omega$, followed by other grammar elements concluding the grammar export. To save time, grammar exportation is delayed until the control loop finds the most descriptive grammar based on user specifications.

\section{Results and Applications}
Our framework is implemented in C++ using Qt, OpenGL, OpenCV, PCL, and dlib on an Intel i7 Desktop PC with a NVIDIA GTX 680 graphics card. We used publicly available 3D databases (e.g., Trimble Sketchup, \cite{lafarge2013}, \cite{Laserscan}) and some manually created models. We applied our guided proceduralization system to polygonal, point-based, and textured models of varying scales, from simple buildings (e.g., 2000 polygons) to large cities (e.g., 4000 buildings) and laser scans (e.g., 2M points), including over 70 individual buildings and 3 large cities. We demonstrate the applications of our controlled grammars in three areas: completion and reconstruction, architectural modeling, and procedural generation.

\subsection{Analysis}
We show an example analysis of varying grammar values in Figure~\ref{fig:analysis} on a complex model of 23K polygons. We sample and demonstrate a subset of the domain $[\Gamma^*_{min},\Gamma^*_{max}]$ yielding visually recognizable patterns increasing from left to right. Notice that even slight changes in the grammar specification can yield new patterns, not only with different rules, but also with different geometric components and similarity labels. The change in the granularity of the rule elements to define the column rule is shown in the insets. The right-most example shows an extreme case where $\Gamma^*>\Gamma^*_{max}$ which is not desirable and over-segments some non-terminals near the inner walls, visualized as colored triangles. The bottom row also shows a similar sampling of models for $\Gamma^*$ for the rotating building pattern of \textit{Turning Torso} landmark with 11K polygons, each instance (right) is paired with its node distribution (left). We observed that the grammar can capture more details based on the user specification, as seen in the white nodes on the rightmost model.

\subsection{Robustness}
\begin{figure}[h]
	\begin{center}
			\includegraphics[width=1\columnwidth]{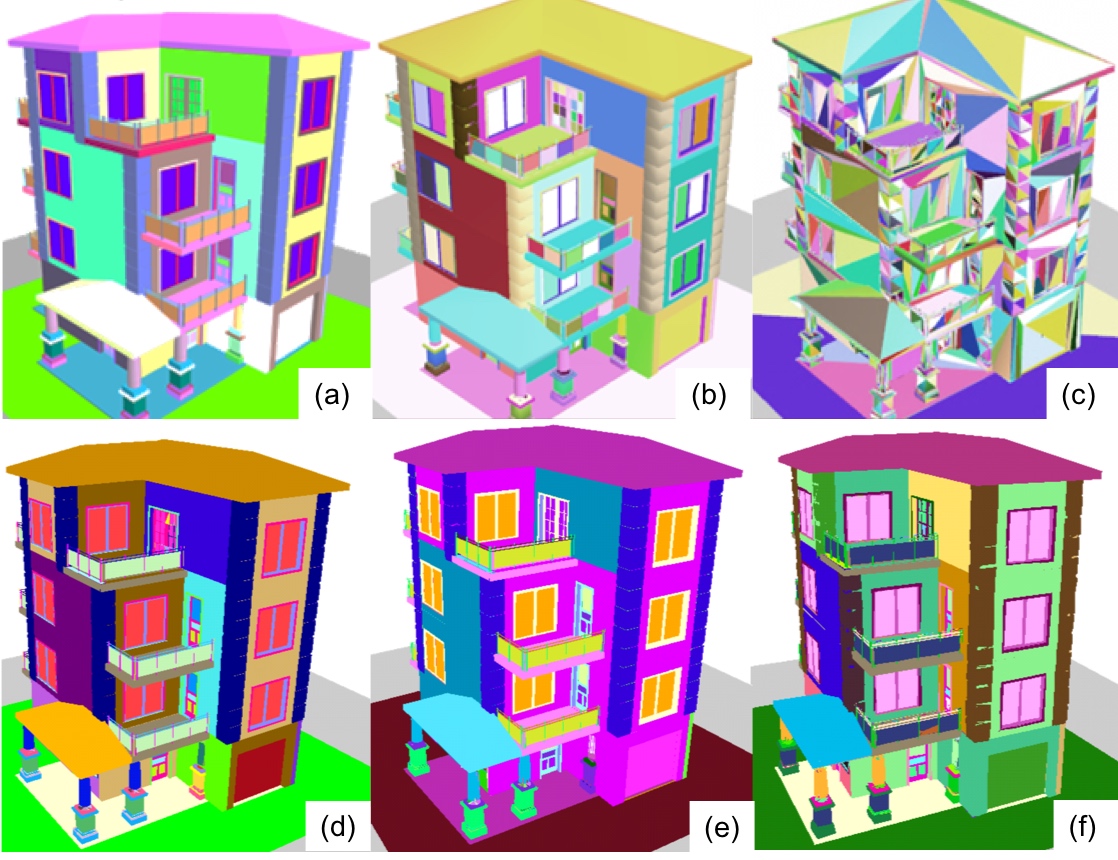}
	\end{center}
	\caption{\textbf{Robustness} comparison between \cite{demir2015} (a-c) and our approach (d-f) against full model vertex displacements (a, d have $0\%$, b,e have $0.1\%$, and c,f have $1\%$). (a,b,c is courtesy of \cite{demir2015})}
	\label{fig:robust}
\end{figure}

We also evaluated the robustness of our approach in case of noisy models. Previous approaches tend to capture perfect repetitions and then convert them to grammars, but they fail to do so when the repetitions slightly differ in geometry and/or in distribution. We observed that guided proceduralization is more robust against such imperfections. We compare our work against work of Demir et al.~\cite{demir2015}, to evaluate the robustness of two approaches on models with different noise levels generated by vertex displacements. Compared to their segmented and labeled model (Figure~\ref{fig:robust}a), $\rho=0.1\%$ full model vertex displacement (meaning that each vertex arbitrarily moves $x\in (0,\rho D]$, where $D$ is the model diagonal) breaks their labeling (Figure~\ref{fig:robust}b), and $\rho=1\%$ vertex displacement breaks the overall algorithm (Figure~\ref{fig:robust}c) (The images are taken from the paper with permission). We input the undisturbed model (Figure~\ref{fig:robust}d) to the forward proceduralization algorithm $P$ and obtained the grammar values $\Gamma$. Then we used the output $\Gamma$ as target grammar values $\Gamma^*$ to control the proceduralization of the disturbed models ($\rho=0.1\%$ in (e) and $\rho=1\%$ in (f)). As shown, our approach was able to find a grammar with the same specification (within $\epsilon$) even for the most noisy model. This also proves the superiority of guided proceduralization over traditional proceduralization in case of noisy input models. Note that the segmentation approaches in both systems are the same, thus the evaluation purely compares the overall frameworks.

\begin{figure}[b!]
	\begin{center}
			\includegraphics[width=1\linewidth]{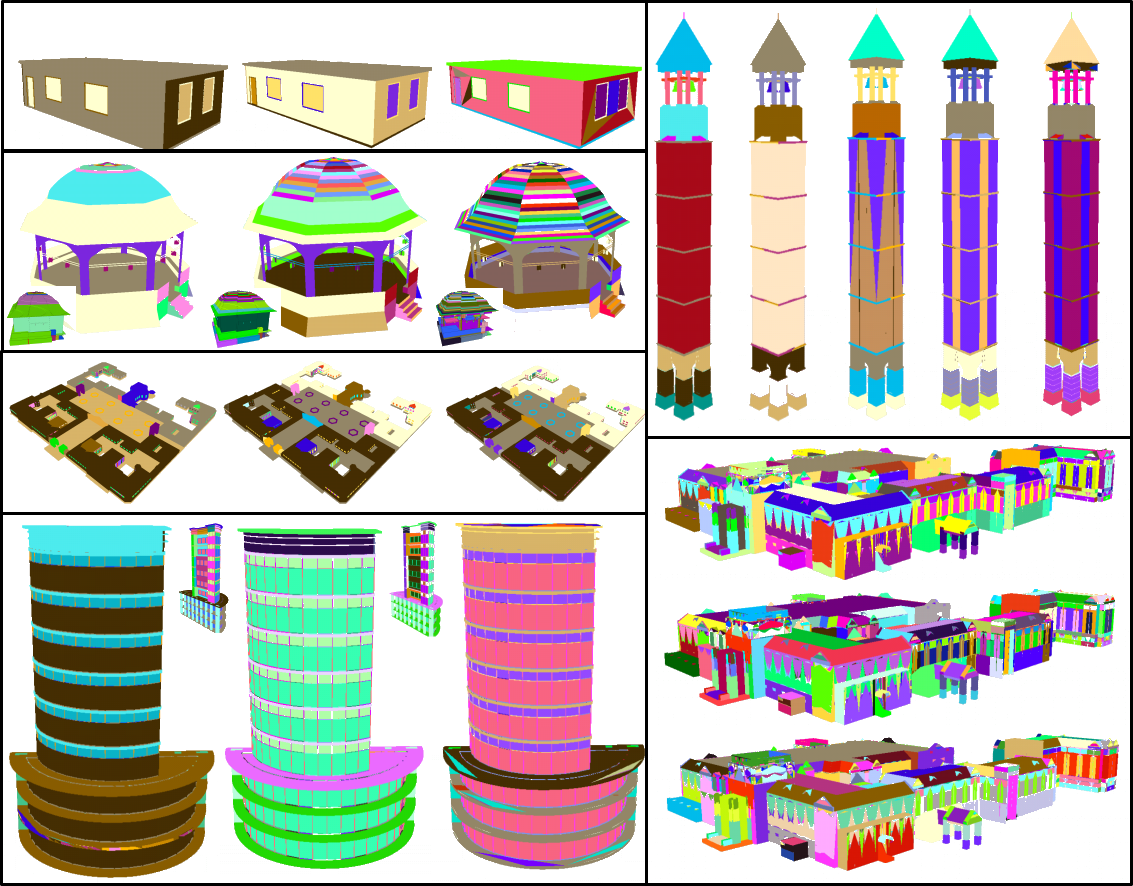}
	\end{center}
	\caption{\textbf{Grammar Variety}. Instead of representing a structure with a unique grammar, we are able to create a family of grammars, for architecturally varying structures. }
	\label{fig:difgram}
\end{figure}

\subsection{Family of Grammars}
We observe that different grammars for different purposes can be obtained using guided proceduralization. In Figure~\ref{fig:difgram}, different rules and different decompositions are colored as repeating patterns on the models. Moreover, we can guide the user towards different use cases by suggesting different grammars per model. This visual suggestion mechanism helps the user to approximate the grammar values $\Gamma^*$ visually, instead of guessing numerically (as explained in Section~\ref{sec:user}). For example, the system suggests horizontal patterns on the left end of the tower body in Figure~\ref{fig:difgram} and vertical patterns on the right end. Similarly, the granularity of captured patterns decreases from left to right for the patio model. We also show that our approach is applicable to models with different structure (i.e., curved buildings), different scale (i.e., neighborhoods), and different complexity (tower with 2K polygons to the union building with 80K polygons).

\subsection{Completion and Architectural Reconstruction}
Although similarities and segmentations are exploited for some reconstruction techniques as in previous methods of Pauly et al.~\cite{pauly2008} and Simon et al.~\cite{simon2012}, the controlled discovery of grammars is not a focus of such papers. In contrast to prior reconstruction methods, we can improve the completion of point clouds before triangulation, without using any priors. We run our controlled proceduralization system on building point clouds, where the user roughly sets $\Gamma^*$ to indicate the desired grammar values. The optimization finds the parameter values $\Theta'_S$ that best segments and labels the point cloud and $\Theta'_D$ that extracts the best grammar. Afterwards, we use all representative instances $v^*_x$ to create consensus models (CM)\cite{demir15ICCV} that are used to improve the quality per instance. The key innovation in using proceduralization for reconstruction emerges from i) exploiting repetitions for completion in a procedural setting, and ii) controlling proceduralization to get rid of the tedious task of semi-automatic segmentation for deciding the segments (as mentioned in Section~\ref{sec:inpdep}). At a first glance, our reconstruction may seem blurrier, however this occurs from the fact that our approach combines all instances of the same terminal into a more complete CM, and overall noise is reduced by edge-aware resampling. The smoothness of the components should not be perceived as blurriness.

\begin{figure}[hb!]
	\begin{center}
			\includegraphics[width=1\columnwidth]{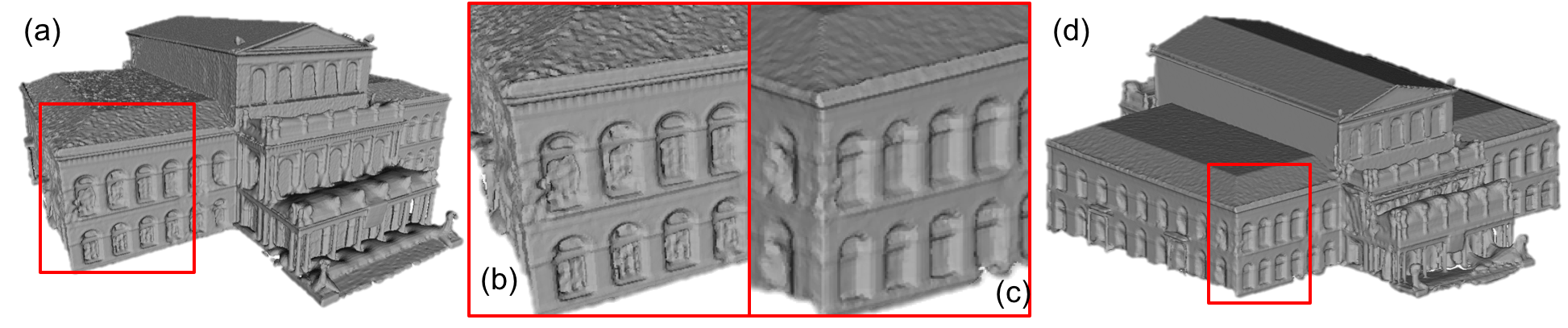}
	\end{center}
	\caption{\textbf{Poisson reconstructions} of (a,b) input and of (c,d) proceduralized point clouds. The identification of representative instances and exploitation of repetitions improve the reconstruction.}
	\label{fig:recon}
\end{figure}

\begin{figure}[ht!]
	\begin{center}
			\includegraphics[width=1\columnwidth]{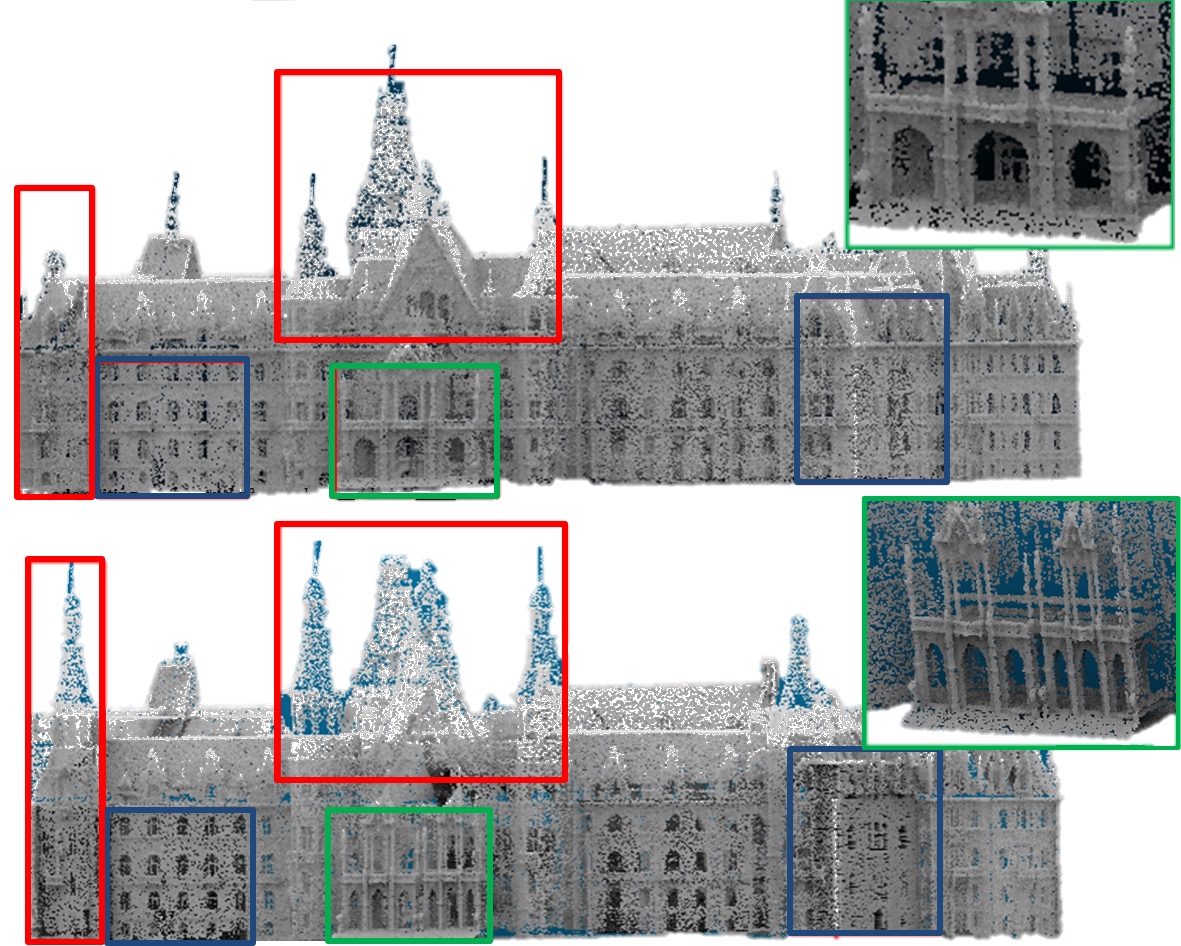}
	\end{center}
	\caption{\textbf{Completion.} Original model (top) and proceduralized and edited model (bottom) are compared for completion and synthesis. Blue boxes focus on improvements for window details, provided by consensus model based proceduralization. Red and green boxes focus on edited parts and their smooth blending, the green box zooming into the entrance of the building.}
	\label{fig:cm}
\end{figure}

Figure~\ref{fig:cm} compares an input point cloud of 1.2M points vs. its proceduralized and edited version. Blue boxes emphasize the regions improved by only consensus models, prior to any editing sessions. Red boxes emphasize the parts after both automatic completion and procedural editing sessions, and how the blending is seamless. The green inset shows an extreme case where the modeler changes the structure is completely after the completion, but the underlying procedural engine is still able to preserve the style. Figure~\ref{fig:recon} also compares a model with 2M points with its proceduralized version, after reconstruction of the point clouds. As shown in the insets, controlled proceduralization as a pre-construction step produces more complete models. The completion is evaluated by comparing the amount of additional points in the new model, after CM models are generated and placed using edge-aware resampling to keep the point density similar  \--- thus additional points most likely means more surfaces  are complete or filled. For example, the completed model in Figure~\ref{fig:recon} has 24\% more points, compared to the original model with a similar local point density in an epsilon.

\subsection{Structure-aware Urban Modeling}\label{sec:edit}
\begin{figure}[h!]
	\begin{center}
			\includegraphics[width=1\columnwidth]{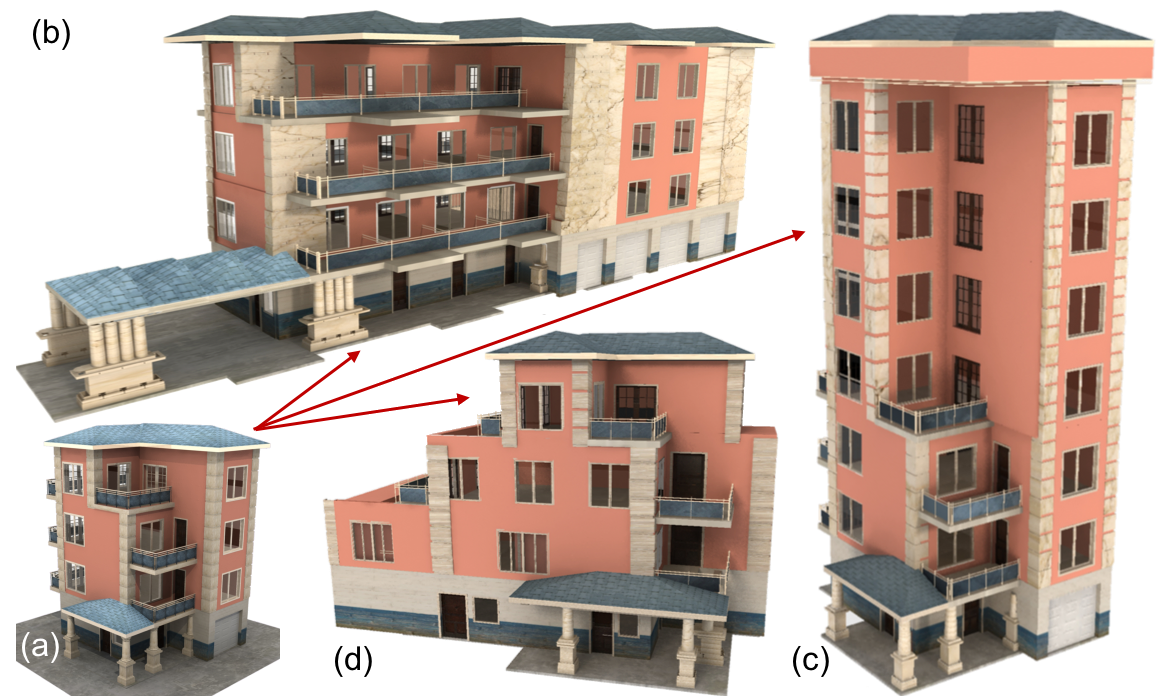}
	\end{center}
	\caption{\textbf{Urban modeling.} Our guided proceduralization allows creation of complex architectures using simple operations, with an average editing time of 15 minutes.}
	\label{fig:edit}
\end{figure}

As mentioned in the introduction, the raw 3D data coming from manual modeling or acquisition systems does not contain any structural or semantic organization. However procedural representations inherently solve that problem and enable faster modeling. We implemented an interactive structure-preserving editing system that uses our procedural engine enhanced with some attachment rules to preserve the adjacencies of grammar elements across rules (similar to~\cite{bokeloh2012}). We convert the instance tree into an instance graph, where the non-tree edges (newly included edges) contain the parameters of spatial adjacencies between not directly related nodes. When the user performs an edit, our framework optimizes the derivation for the propagation of the operation from the edited node to other nodes in the graph, preserving the attachment rules. After the optimization converges, the parameters are re-computed and the model is ready to be exported. Figure~\ref{fig:edit} shows variations of a complex building structure synthesized by our editing system built upon the controlled proceduralization framework emphasizing (b) horizontal patterns, (c) vertical patterns, and (d) mixed patterns. The processing takes less than an hour to process and editing sessions are kept under 5 minutes for this particular model.

Without guided proceduralization, some users were confused about which patterns to use while editing. However, enabling user influence while generating patterns helped users to more intuitively edit the patterns that they defined, and reduced the overall editing time. Another benefit of guided proceduralization over proceduralization is that it reveals some patterns that would never have been discovered by prior proceduralization methods. Consider the last row of Figure~\ref{fig:min} and the last column of Figure~\ref{fig:gram}. The pattern parameters for those non-terminals are highly noisy, but guided proceduralization still captures the essence of double windows. Such cases are especially important in editing systems where the granularity of patterns should follow user's intuition. Our system enables such implicit granularity declarations over previous work converging on the smallest component configuration. A real-world example of this improvement can be observed at the roof of the gazebo in Figure~\ref{fig:difgram}. Previous approaches would have discovered only the right-most configuration with too many rules and terminals (each indicated in a different color), in search of perfect repetitions. In contrast, guided proceduralization enables more compact configurations by auto-adjusting the granularity and rule/terminal similarity following the user guidance (instances are colored with the same color), adding the flexibility to fit imperfect repetitions into rules.

\subsection{Procedural Content Generation}
\begin{figure}[ht!]
	\begin{center}
			\includegraphics[width=1\linewidth]{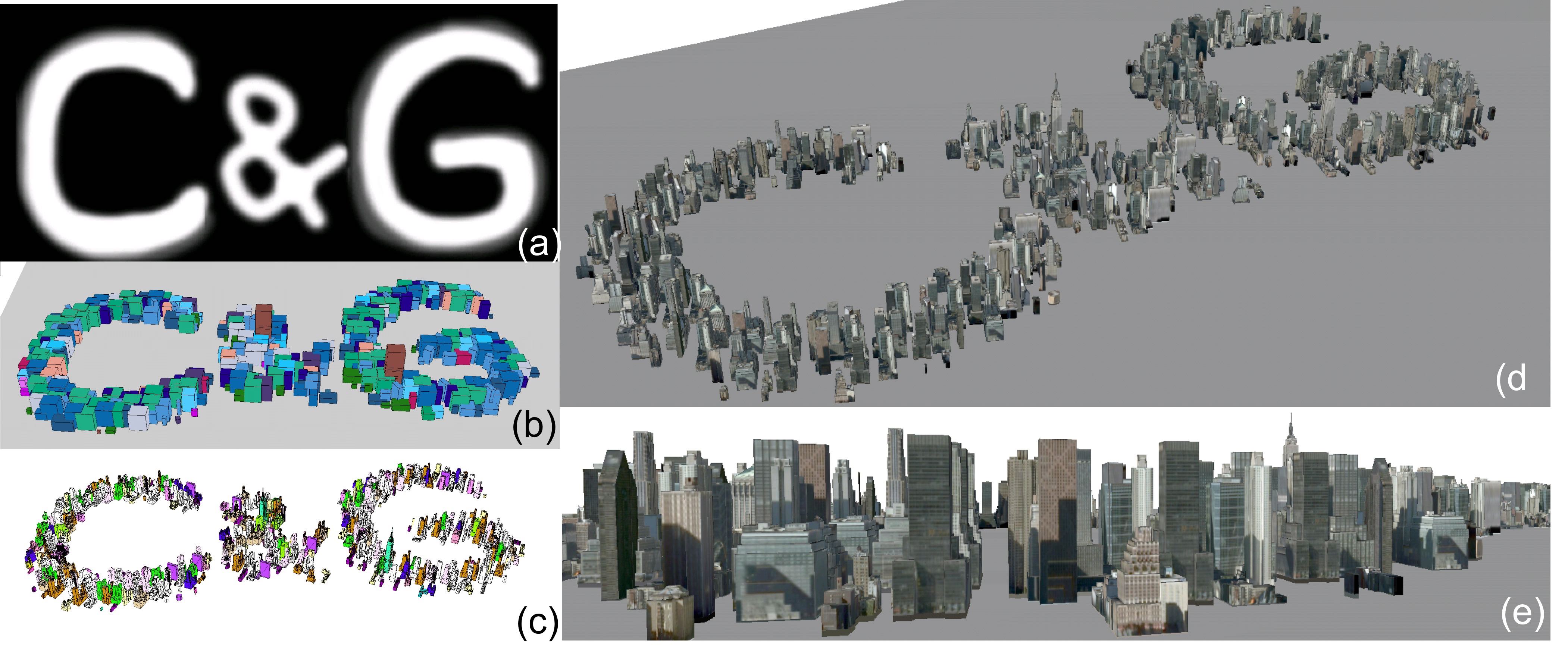}
	\end{center}
	\caption{\textbf{Procedural Generation.} (a) Input map $\omega=$``C\&G'', followed by the rendered model in (b) color-coded tree nodes, (c) labeled buildings, (d) textured buildings, and (e) street view.}
	\label{fig:cg}
\end{figure}

Finally, we use our controlled proceduralization framework to create procedural models for large content creation, which is used for compression and efficient rendering. Previous approaches (e.g., \cite{wonka2003, vanegas2012, ritchie2015, demir2014, talton2011}) either need a grammar for inverse procedural modeling, or use thresholds to create derivation trees. However our approach keeps the details of the procedural model under the hood and enables content generation based only on a geometric model, and optionally grammar specification values. We used our controlled proceduralization framework to convert 4K buildings in the $180km^2$ metropolitan area of New York into a procedural representation, achieving 95\% compression in size (reducing from 1.7M polygons to175K polygons, and from 1060MB of textures to 49MB) . We show a new city (called ``CG York'') generated using the extracted grammar in Figure~\ref{fig:cg}. We also demonstrate input map, nodes of the derivation tree, color-coded components, and a street view from our new city.

\section{Discussions}
\subsection{Expressiveness}\label{sec:expressive}
At first glance, one can claim that finding the most effective grammar (i.e., compact)
is not possible, because finding the minimal description $|d(.)|$ covering all instances of all possible grammars requires computing their Kolmogorov complexity (i.e., $K(.)$), which is not a computable function. However, in our case i) grammars can be regarded as compressors that implicitly eliminate some redundant information and simplify the description, and ii) we do not seek a minimal description (i.e., $|d(.)|=K(.)$) but a description that approximates our target grammar specification values ($|d(.)|\approx K(.)$). We incorporate user specified $\Gamma^*$ to decide on the level of simplification and to produce the most useful grammar based-on-need, as shown in different decompositions and rules of Figure~\ref{fig:analysis}. For example, if we were to look for the minimal description of ``abcabababab'', we would have to find a description including ``c''. But if the user indicates that single occurrences are noise, or the label ``c'' can be simplified, or wants four instances, then an easy and efficient description would be ``(ab)*5''. In Figure~\ref{fig:analysis}, such user specification is incorporated as different $\Gamma^*$ values, and the results indicate how much variance is tolerated within each terminal group (similarity noise) and within each rule (like the "ab" example). 

\subsection{Guided Proceduralization vs. Controlled IPM}
One option that some exploration algorithms in inverse procedural modeling use is providing the user with some example derivations and letting them guide the generation. In prior work (e.g., \cite{yumer2015,talton2011,ritchie2015,nishida2016}) this approach assumed a known grammar and a discretization of the derivation space. In our case for proceduralization, we let the user specify desired grammar values ($\Gamma^*$) beforehand and perform an optimization. The distinction of those processes are more clear when visualized: Bottom row of Figure~\ref{fig:robust} shows guided proceduralization output (the desired grammar), and Figure~\ref{fig:edit} shows controlled IPM outputs (the desired instances of a grammar). For clarification, we do not input a grammar (e.g. $G=E(\Gamma)$), but we input perceptible grammar values ($\Gamma^*$), i.e., the alphabet size, repetition per rule, etc. (Section~\ref{sec:gramval}).

\subsection{Optimization Space}
The optimization search space $\mathcal{P}(\Theta)$ has variables of geometric and topological similarity as discussed in definitions of $\Theta_S$ (Section~\ref{sec:thetas}) and $\Theta_D$ (Section~\ref{sec:thetad}). We set appropriate bounds for the search space parameters and run multiple iterations of the proceduralization process until we converge on the desired grammar specification values. Also, instead of a brute force exploration, we use BOBYQA algorithm~\cite{powell09} to converge faster in this well-bounded space without derivatives. This reduces the execution time of the optimization, from 1-1.5 days to 4-5 hours on a single machine with 8 CPUs, for a model of ~20K triangles. The optimization is also flexible enough to include other parameters $\Theta$ and specifications $\Gamma$ if needed. We implemented an import/export mechanism for our intermediate steps. Thus, the current grammar can be exported at any stage of the optimization, which provides flexibility to start from near-optimum grammars for faster convergence. Also, the interactive user interface allows the underlying system to use the grammar from the last step of the optimization whenever new target values are set by the user, instead of starting from scratch.

Although our prototype optimization may take a couple of hours, converting the models into the desired procedural representation is a one-time offline operation that does not consume any human labour. In addition, the procedural editing engine is completely online, enabling the user synthesize many complex models in minutes, following the procedural control that they defined. It is also possible to make the offline part more efficient. To converge faster, one can structure $\Theta$ to include differentiable features, or to create a convex optimization space, and then change the algorithm to a gradient-descent variation. Carrying the computation to GPUs and introducing parallelization for the optimization is also a suggested option. However in our experiments, we targeted for the most flexible setup to support all possible features.

\subsection{User Interaction}\label{sec:user}
We have implemented two means of defining target grammar values $\Gamma^*$ for users. The first one presents some GUI elements (sliders and input boxes) for defining the grammar elements defined at the end of Section 4.2.1. Users can set these relative values using the GUI, and the system optimizes for the grammar that best satisfies these values. The second setting visually provides suggestions as a variety of grammars where elements of each grammar choice are color-coded on the re-generated input model (Figure~\ref{fig:difgram}). We achieve that by sampling the multi-dimensional grammar value space $[\Gamma_{m_{min}},\Gamma_{m_{max}}]$ and filtering the results that converge to significantly different grammars. Then, we re-generate the input model using each of the candidate grammars, and demonstrate these instances of candidate grammars. The user can then select the best grammar by choosing the instance that visually contains the patterns and components that are applicable to her use case.

After the grammar is generated, the procedural editing engine provides operations like push-pull, copy-paste, and replicate-join. All of those operations are conducted visually on the model via mouse/keyboard interactions. The user can copy terminals and non-terminals by clicking on them, which are adaptively pasted into their new nodes to fill its volume. Replicate and join operations change the application counts of patterns by simple strokes on the terminals and non-terminals. Finally, the user can pull a model and the procedural engine resizes the model as explained in Section~\ref{sec:edit}. Note that this operation is different than just changing pattern parameters, because the adjacency graph of the nodes are preserved during editing.

\section{Conclusion}

In summary, our research shows that extending proceduralization frameworks to include user guidance discovers procedural representations for existing models so as to improve reconstruction and modeling. As the graphics and vision communities converge in automatic asset creation in almost all domains, many fascinating problems await to be solved by procedural approaches instead of manual and semi-automatic organization of reconstructed geometry.

Looking forward, we have identified the following items of future work: (i) using a neural network model to approximate function $f(\Theta')\rightarrow \Gamma^*$ for better performance (to replace hours of optimization  with minutes of prediction), (ii) bringing our controlled proceduralization framework to other domains, and (iii) proposing evaluation tools for output grammars to assess their generative capacity.

\section*{Acknowledgments}
This research was funded in part by National Science Foundation grants ``CDS\&E:STRONG Cities - Simulation Technologies for the Realization of Next Generation Cities''€ CBET 1250232 and ``€œCGV: Medium: Collaborative Research: A Heterogeneous Inference Framework for 3D Modeling and Rendering of Sites''€ IIS 1302172. We also thank Matt Sackley for helping with some of the rendered building models, and Clark Cory for the modeling support. We would also like to acknowledge the open and free, crowd-sourced and academic 3D databases for enabling our research to be conducted on hundreds of different models and formats.

\bibliographystyle{cag-num-names}
\bibliography{egbib}

\end{document}